\documentclass[aps,preprint,showpacs,superscriptaddress]{revtex4}
\usepackage{amssymb}
\usepackage{bm}
\usepackage{graphicx}
\usepackage{amsmath}
\usepackage{url}
\usepackage{multirow}
\usepackage{longtable}

\newtheorem{remark}{Remark}
\newcommand{\Tr}{{\rm Tr}}

\def \b{\begin{equation}}
\def \e{\end{equation}}
\def \ba{\begin{array}}
\def \ea{\end{array}}
\def \be{\begin{eqnarray}}
\def \ee{\end{eqnarray}}
\def \n{\nonumber}
\def \l{\left}
\def \r{\right}

\begin{document}
\title
{Laser and Diffusion Driven Optimal Discrimination of Similar Quantum Systems in Resonator.}

\author{\firstname{K. A.}~\surname{Lyakhov}}
\email[E-mail: ]{lyakhov2000@yahoo.com}
\affiliation{Department of Mathematical Methods for Quantum Technologies, Steklov Mathematical Institute of Russian Academy of Sciences, Gubkina str. 8, Moscow 119991, Russia}
	
\author{\firstname{A. N.}~\surname{Pechen}}
\email[E-mail: ]{apechen@gmail.com}
\affiliation{Department of Mathematical Methods for Quantum Technologies, Steklov Mathematical Institute of Russian Academy of Sciences, Gubkina str. 8, Moscow 119991, Russia}
\affiliation{National University of Science and Technology MISIS, Leninskii Avenue 4, Moscow 119049, Russia}

\begin{abstract}
In this work, a method for solving the problem of efficient population transfer from the ground to some excited state by available technical means (varying resonator length) is proposed. We consider a mixture of similar quantum systems distributed in a stationary gas flow in the resonator with variable resonator length, which implements tailored laser field close to the optimal one. In difference from previous works, in this work piezoelectric transducer (PZT; or actuator) is used as mean of control to manipulate variable resonator length. The external actions are optimized to selectively prepare different stationary states of different species which are then separated using their diffusion with rates which differ due to different masses. This system provides an example of the general problem of optimizing states of distributed systems with stationary gas flow with diffusion. 
\end{abstract}


	
\maketitle

\section{Introduction}
	
Optimized separation of mixtures of similar quantum systems, such as e.g., isomers, isotopes, or flavins, is an important problem in quantum control~\cite{Ref1,Ref2,Ref3,Ref4,Lee77,VdBer85,Lym2005,Eer2005,Lya2017,Lya2018,Lya2020_a,Lya2021}.  For separation of these systems various methods such as optimal dynamic discrimination method~\cite{Ref1,Ref2,Ref3,Ref4} and the laser assisted retardation of condensation method~\cite{Lee77,VdBer85,Lym2005,Eer2005,Lya2017,Lya2018,Lya2020_a,Lya2021} can be employed. 
	
The problem of discrimination of quantum systems immersed as a supersonic gas flow in a resonator and excited by tailored laser pulse is being addressed in this study.  Selective action of the laser beam relies on resonant absorption of laser photons by target molecules, that prevents their condensation which otherwise would naturally occur. Thus, the mass difference between light (excited) and heavy (not excited) species is  artificially induced by applying selectively acting laser beam radiation to over-cooled gas flow. In supersonic gas flow light species diffuse to gas flow periphery faster than heavier ones. Hence, heavy and light species, comprising rim and core gas flows, respectively, due to their different spatial distributions can be separated by the skimmer blade at the end of the gas flow. 
	
Since in over-cooled gas flow photo-absorption spectra are very narrow, the proper choice of laser emission line seems crucial. For example, it is known that even at exact resonance (at continuous wave irradiation) only $\approx14$\% of the ground state population of $H_2O$ molecules can be transferred to some specific excited state~\cite{Jak90}. In order to achieve $\approx99$\% of the ground state population transfer, one needs laser pulses with duration $t_p>10{\rm ns}$ and with specially tailored shape. At low values of static pressure within the over-cooled gas flow which is composed of small target gas molecules this requirement can be met, but in this case in order to achieve high enough production rate it is necessary to use multi-nozzle setup which will increase overall facility size substantially. This problem can be solved by using moderate values of pressure in the gas flow. In order to increase the enrichment factor, which grows with the excitation rate, one needs to be able to control laser field evolution in such a way that excitation rate is much higher than excitation loss at elevated static pressure in the gas flow. Laser field choice should also minimize the effect of intramolecular vibrational redistribution for large molecules. 

In contrast to previous studies~\cite{Lya2017,Lya2018,Lya2020_a,Lya2021}, in this work an appropriate laser field evolution is modulated by laser resonator length (its active partition) by applying periodic voltage to piezoelectric transducer which transforms applied voltage into mechanical deformation (PZT, sometimes called as piezoelectric actuator to distinguish from piezoelectric sensors~\cite{ZHANG201073,Somiya2003}), which is used as a mean of control. The control field may generally be spatially dependent along width of the irradiation cell and the gas flow is spatially distributed along the length of the cell, thereby making the system as controlled distributed system with diffusion. We formulate the control problem as maximization of excitation probability of some component of the mixture and minimize probabilities of excitation of all other components. The objective is considered as a functional of the PZT deformation. We find explicit expression for the gradient of the objective as a functional of the applied electromagnetic field which forms a basis for gradient ascent optimization which we develop for this problem.
	
Structure of the paper is the following. In Sec.~\ref{Sec:General}, general formulation of the method is proposed along with model of technical implementation of the considered model setup. In Sec.~\ref{Sec:Gradient}, gradient ascent search method for this model is developed, including explicit expression for the gradient of the objective functional. In Sec.~\ref{Sec:Transport}, the transport model is presented. Section~\ref{Sec:Intracavity} provides model for the intracavity field evolution. Sections~\ref{Sec:Excitation} and~\ref{Sec:Crosssection} contain description of the excitation rate and photo-absorption cross section, respectively. Section~\ref{Sec:Conclusion} summarizes the results.
	
\section{General formulation of the method}\label{Sec:General}
	
\begin{figure}[ht]
\includegraphics[width=.9\linewidth]{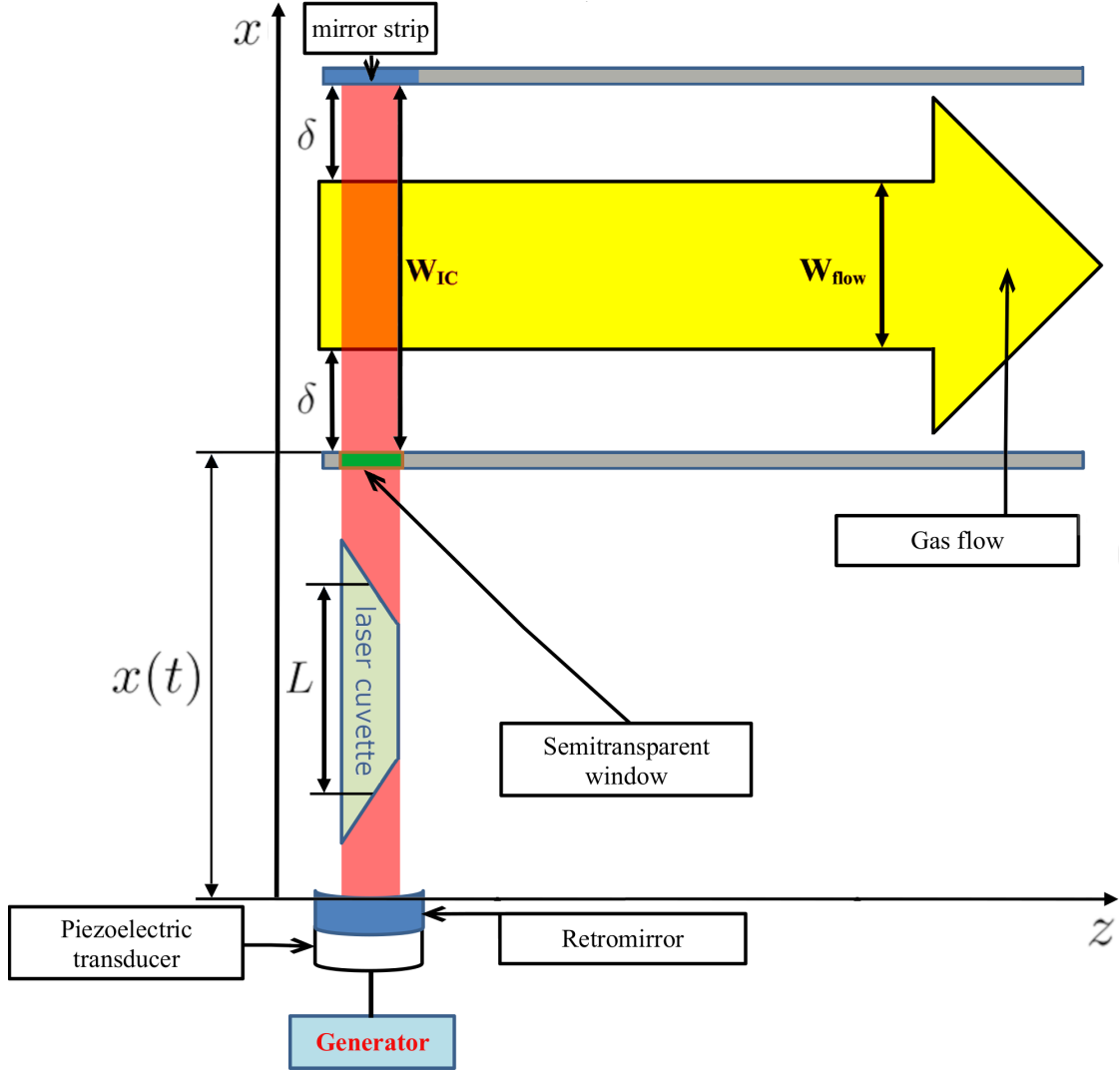}
\caption{\label{design} Schematic of the experimental setup that can be used for optimal shaping of control laser pulses by applying periodic voltage to piezo-electric transducer (PZT). Modulated by PZT laser pulse enters into the cell and selectively excited only one component of the mixture in the gas flow. Parameters of the PZT  (amplitude and frequency, or its more general time dependent deformation) are optimized to maximize the probability of excitation of this component while minimizing probability of excitation of other components. Then the components are separated via diffusion when moving in the gas flow to the right in the rest of the cell.}
\end{figure}
	
Technical implementation of the proposed laser field control in the absorbing partition of laser resonator is shown in Fig.~\ref{design}. The control mean is PZT (or actuator) deformation $u(t)\in{\cal U}$ which is induced by applied voltage and is used to modulate electromagnetic field inside the cell as described in Sec.~\ref{Sec:Intracavity}. Here $\cal U$ is class of  PZT deformations allowed in given experimental setup.
	
In general formulation, we consider a gas phase mixture of $K$ various types of quantum systems with similar spectra, which enters the irradiation cell from left side on the picture (the wide yellow arrow shows the gas flow). In this mixture, each quantum system of type $k$ has free Hamiltonian $\hat H^0_k$ and it interacts with electromagnetic field via its dipole moment $\hat\mu_k$. PZT deformation $u(t)$ is a control action which is used to modulate electromagnetic field inside the cell. Hence electromagnetic field inside the cell is a functional of $u(t)$, ${\cal E}(x,t)={\cal E}_u(x,t)$. 
	
In the simplest approximation, we can consider averaged over gas flow width $W_{\rm flow}$ electromagnetic field ${\cal E}_u(t)$. This electromagnetic field determines evolution of the wave function $|\Psi_k^{{\cal E}}(t)\rangle$ of the $k$-th component molecule in the cell according to Schr\"odinger equation
\b
i\frac{d|\Psi_k^{{\cal E}}(t)\rangle}{dt}=\Bigl(\hat H^0_k+\hat\mu_k {\cal E}(t)\Bigr)|\Psi_k^{{\cal E}}(t)\rangle
\e
Generalization to spatially distributed electromagnetic field is straightforward. Dipole moment and electric field are vector valued quantities, $\hat{\vec\mu}_k$ and $\vec{\cal E}(t)$, but we omit vector signs for brevity. The control goal is to maximally excite $i$th target component and minimally excite all other components of $K$-component mixture. Then the subsequent spatial separation can be made at the skimmer blade. For simplicity one can assume that initially all molecules are in the ground state; generalization is straightforward.
	
Let ${\cal P}_k$ be projector onto the excited space of a $k$ component quantum system, 
\b
{\cal P}_k(t)=\sum\limits_{\textrm{excited states}}|\Phi_{k,i}\rangle\langle \Phi_{k,i}|,
\e
where $|\Phi_{k,i}\rangle$ is an excited state (basis state in the absence of laser field) of the $k$th molecule. Then probability of excitation of the $k$th type molecules at time $t$ is 
\b 
P_k^{\rm ex}=\langle \Psi_k^{{\cal E}}(t)|{\cal P}_k|\Psi_k^{{\cal E}}(t)\rangle
\e
	
The control goal can be formulated as maximization of the probability to excite the $j$th component, after a single laser pulse of duration $\tau_p$, while minimizing probabilities of excitation of all other components. This control goal can be represented as maximization of the control objective functional 
\b
{\cal F}(u)= P^{\rm ex}_j(\tau_p)-\sum\limits_{k\neq j}^K\alpha_kP^{\rm ex}_k(\tau_p)\to\max
\e 
Here $\alpha_k>0$ is weight which regulates the importance of not exciting of all other $K-1$ components (in a simplest situation all $\alpha_k$ are equal). Negative terms, corresponding to $k\neq j$, indicate that we want to minimize excitation of these remained components.

\begin{remark}
One could alternatively consider the objective as a functional of the electromagnetic field ${\cal E}(t)$, ${\cal F}({\cal E})$, and first find an optimal shape of control field ${\cal E}_*$. Then one could find unknown form $u(t)$ of the PZT deformation as a solution of the following optimization problem 
\b 
\min\limits_{u\in{\cal U}}\l\{\|{\cal E}_*- E\|_{C_{[0,\tau_p]}}\right\},
\e 
where ${\cal E}_*$ and $E$ are optimal laser electric field to be found and its technical implementation induced by some $u$, respectively; minimization has to be performed allowed shape of $u(t)$ which should be determined for a given experimental availability. However, in this case it might happen that the set of electromagnetic field generated by allowed $u\in{\cal U}$ is too far from the optimal shape ${\cal E}_*$, and finding closest ${\cal E}$  induced by a PZT deformation $u$ is not the best choice: there may exist another $u_*$ which generates $E_{u_*}$ which has higher objective value. Therefore we consider directly the functional ${\cal F}(u)$ as the control objective.
\end{remark}
	
\section{Gradient ascent search}\label{Sec:Gradient}
To find a local maximum of $\cal F$, we propose to use gradient based search. Its advantage for this problem is that gradient of the objective to some extent can be computed exactly. Indeed, variation of $\cal F$ over $u$ can be found using the chain rule
\b
{\rm grad} {\cal F}=\frac{\delta {\cal F}}{\delta u}=\frac{\delta {\cal F}}{\delta {\cal E}}\frac{\delta \cal E}{\delta u}
\e
In the variation ${\delta F}/{\delta {\cal E}}$, the objective and transition probabilities are considered as functionals not of $u$, but of directly laser field $\cal E$. The variation ${\delta F}/{\delta {\cal E}}$ can be found for example using famous Gradient Ascent Pulse Engineering (GRAPE) approach~\cite{KhanejaJMR2005.2.296} commonly used to find NMR optimal pulse shapes, where the time dependent Schr\"odinger equation for the target system wave function is naturally included in the expression for the gradient of the objective functional. For deriving explicit expressions for the variation ${\delta F}/{\delta {\cal E}}$ one can use general Eq.~(1) from Supplementary material in~\cite{PechenPRL2011}, by setting in this equation $O={\cal P}_j-\sum\limits_{k\ne j}\alpha_k{\cal P}_k$ and $T=\tau_p$. Computing variation ${\delta \cal E}/{\delta u}$ of the intracavity laser field over PZT deformation $u$ can be done using results of Sec.~\ref{Sec:Intracavity}.

Denoting $V^{\cal E}_k(t)=U_{k,\cal E}^\dagger(t)\mu_k U_{k,\cal E}(t)$, where $U_{k,\cal E}(t)$ is the evolution operator of $k$th component of $K$-component quantum system in the presence of laser field $\cal E$, we may obtain expressions for the variation of the excitation probability:
\b\label{gradient}
\frac{\delta P_k^{\rm ex}}{\delta {\cal E}(t)}=-i\Tr\left\{ \Bigl[\rho^k_0,{\cal P}_k(\tau_p)\Bigr] V^{\cal E}_k(t)\right\}
\e
where $\rho_0^k$ is the projector on the initial state of $k$th component (its initial density matrix in the general case), ${\cal P}_k(\tau_p)=U^\dagger_{k,\cal E}(\tau_p){\cal P}_kU_{k,\cal E}(\tau_p)$, and square brackets $[\cdot,\cdot]$ denote commutator of two operators. Hence,
\b
\frac{\delta\cal F}{\delta {\cal E}(t)}=-i\Tr\left\{ [\rho^j_0,P_j(\tau_p)] V^{\cal E}_j(t)\right\}+i\sum\limits_{k\neq j}^K \alpha_k \Tr\left\{ \Bigr[\rho^k_0,{\cal P}_k(\tau_p)\Bigr] V^{\cal E}_k(t)\right\}
\e

Then the search for a locally optimal control can be performed via step by step update of the laser pulse shape 
\b 
{\cal E}_{i+1}={\cal E}_i+\epsilon {\rm grad}{\cal F}
\e
Step size $\epsilon$ can be variable in general. Control field can be considered as a 3D vector if necessary. 

Of course, gradient based search generally will converge not necessarily to global but possibly to a local maximum of the control objective. If local maxima exist, then one has to run gradient search several times  starting from various initial conditions and select the best outcome. The question whether local maxima for quantum control objective functionals exist or not is important in quantum control~\cite{RabitzScience2004}. Absence of local maxima was proved for two-level quantum systems~\cite{PechenPRA2012,VolkovJPA2021} and for quantum systems with infinite-dimensional Hilbert space (control of transmission)~\cite{PechenCJC2014}, while general case remains mostly unsolved.  Moreover, trapping behaviour was found for some multi-level quantum systems~\cite{PechenPRL2011}. Absence of traps in the kinematic picture was also established for control of open quantum systems, where for the first time complex Stiefel manifold description of open quantum system dynamics was proposed and gradient optimization theory was developed~\cite{PechenJPA2008,OzaJPA2009,PechenPRA2010.82.030101}, as well as for classical systems~\cite{PechenEPJ2010}. Machine learning~\cite{MorzhinLJM2021} and stochastic optimization methods such as dual annealing can also be used for optimization of transition probabilities and state-to-state transitions~\cite{MorzhinLJM2021}.

Lagrange multipliers method may also been employed which according to~\cite{Jak90} allows to save more than two orders of magnitude of computational time comparing to direct propagation of equivalent wave packets. 

In general, a feedback between the output signal and the PZT with learning control~\cite{Ref4} and possible nonlinear reaction term can be used to better manipulate the resonator if some relevant parameters of the components (e.g., free Hamiltonians or dipole moments) are unknown. In this case the problem will be an example of optimizing stationary states of a distributed system with diffusion and non-linear reaction term.

\section{The Transport Model}\label{Sec:Transport}
	
Below we consider as an example finding optimal laser pulse shape delivering maximum of enrichment factor at fixed all other macroscopic parameters such as gas flow pressure $P$ and temperature $T$, target gas molar fraction $\mu$, etc. Enrichment factor can be introduced as 	\b\label{enrich0}\beta=\frac{^{i}Q_{\rm esc}/Q_{\rm esc}}{^{i}Q_{\rm feed}/Q_{\rm feed}},\e  where $^{i}Q_{\rm feed}/Q_{\rm feed}$ and $^{i}Q_{\rm esc}/Q_{\rm esc}$ are relative abundances of $i$th sort monomers in the escaped and feed gas flows respectively. According to \cite{Eer2005}, enrichment factor increases with increase of molar fraction of selectively excited monomers. The latter increases with laser excitation rate $k_{\rm A}$. Therefore, the goal is to find optimal laser pulse shape to maximize photo-absorption rate. 
	
Consider the situation when laser excitation of $i$th sort of target gas molecules in rarefied over-cooled gas flow can be modeled by four characteristic groups, specified by their molar fractions, denoted by $f_{i \rm m}$ for monomers, $f_{i\rm *}$ for	excited monomers, $f_{i!}$ for epithermals, and $f_{i \rm d}$ for dimers. Then molecule excitation dynamics by selectively tuned laser field in supersonic flow can be described by the transport model proposed in~\cite{Eer2005}. The material balance equation reads as 
\b 
f_{i\rm m}+f_{i*}+f_{i!}+f_{i \rm d}=1.\n
\e 
Evolution of the molar fractions is described by the following system of equations
\b\l\{\ba[c]{l}
	\frac{df_{i*}}{dt}=k_{\rm A}(t)f_{i\rm m}-f_{i*}\{(1-e_*)(k_{\rm df}+k_{\rm VV}+k_{\rm VT}+k_{\rm se})+e_*k_{\rm W}\},\\\frac{df_{i!}}{dt}=(1-e_*)(k_{\rm df}+k_{\rm VT})f_{i*}-\{(1-e_1)k_{\rm th}+e_1k_{\rm W1}\}f_{i!},\\
	\frac{df_{i\rm d}}{dt}=k_{\rm df}f_{i\rm m}-k_{\rm dd}f_{i\rm d}.
	\ea
	\r.\label{model_0}
	\e
	Here, $k_{\rm df}$ and $k_{\rm dd}$ are the dimer formation  and
	dissociation rates, $k_{\rm VT}$ and $k_{\rm VV}$ are the
	vibration-translational and vibration-vibrational photon spontaneous
	emission rates, $k_{\rm se}$ is the photon spontaneous emission
	rate, $k_{\rm th}$ is the epithermals thermalization rate, $k_{\rm
		W1}$ is the "gas flow wall" escape rate for epithermals, $e_*$ and $e_1$ are
	the to-the-wall survival probabilities for excited monomers and
	epithermals respectively. Explicit formulas for transport coefficients and their derivation can be found in \cite{Eer2001}.

\section{Intracavity Laser Field Evolution}\label{Sec:Intracavity}
	
Laser pulse wavelength $\lambda_{\rm las}(t)$ should meet the following resonance condition: 
\b \label{Eq:Wavelength}
	N\frac{\lambda_{\rm las}(t)}{2}=nW_{IC}+x(t),\qquad N\in\mathbb N
\e 
where $n$ is index of refractivity of the gas in the irradiation cell (IC) at given pressure, which we here consider to be constant, $W_{IC}$ is the irradiation cell width, and
	\b 
	x(t)=x_0+u(t),
	\e 
where $u(t)$ is PZT deformation induced by applied voltage, $|u(t)|\ll x_0$. In the cell, regions of size $\delta$ are filled with carrier gas, while $W_{\rm flow}$ are filled with mixture of carrier and target gases. Thus they may have (slightly) different refraction indices. This spatial dependence can be taken into account in~(\ref{Eq:Wavelength}), if necessary. Quantity  $x_0$ is the distance between PZT surface placement without applied voltage and upper surface of the semitransparent mirror (strictly speaking $x_0$ is the corresponding to this distance optical path length, which takes into account possible variability of indexes of refractivity along the path), which fulfills to resonance condition, corresponding to undisturbed laser emission wavelength $\lambda_{\rm las}^0$: 
	\b 
	x_0+nW_{IC}=N\frac{\lambda_{\rm las}^0}{2}.
	\e 
In the particular case of periodic deformation $u(t)=A\cos\Omega t$. More general shape of the PZT deformation can be used when technically available.
	Thus, in general case
	\b 
	N\frac{\lambda_{\rm las}(t)-\lambda_{\rm las}^0}{2}=u(t),\qquad N\in\mathbb N.
	\e
	
	Time evolution of the laser intensity $I(x,t)$ inside the passive partition of the laser resonator can be evaluated by:
	\b\label{intens}
	I(x,t)=\frac{c\epsilon_0n|\vec E(x,t)|^2}2.
	\e
	Here $c$ is the speed of light and $\epsilon_0$ is vacuum permittivity.
	
	Since the laser beam is confined between two parallel mirrors, which form a multi-pass resonant optical cavity, the electric field at some fixed time $t$ at point $x$ is a superposition of the electric fields $E^{(j)}(x,t)$ from the previously emitted laser pulses over the total time interval elapsed $t$:
	\b
	\vec E(x,t)=\sum\limits_{j=0}^{N_p(t)}\vec E^{(j)}(x,t).
	\e
	Here $N_p(t)$ is the number of emitted laser pulses; the first term $\vec E^{(0)}(x,t)$ in the summation corresponds to the current laser pulse. 
	
	The number of the laser pulses $N_p(t)$ can be calculated as the upper bound of the ratio $t/T_p$, where $T_p$ is the length of the laser pulse train. 
	
	Following Ref.~\cite{Kuiz70}, we assume Gaussian envelope for intensities. Thus, total electric field strength of mode-locked laser pulses is:
	\b\label{field_str}
	\vec E^{(j)}(x,t)=\vec E_0\Phi^{(j)}(x,t),
	\e
	where 
	\begin{eqnarray}
		\Phi^{(j)}(x,t)&=&\frac1{2\pi}\int\limits_{-\infty}^\infty d\omega Q_{\rm las}(\omega,x,t){\cal F}(\omega,t) e^{i\omega t_{j}},\\
		{\cal F}(\omega,t)&=&\frac{4\sqrt{\pi\ln2}}{\Delta\omega_p(t)}\exp\l\{-4\ln 2\l[\frac{\omega-\omega_{{\rm las}}(t)}{\Delta\omega_p(t)}\right]^2\right\},\\
		t_{j}&=&t-jT_p-\tau_r^{(j)}
	\end{eqnarray}
	Here, $\omega_{{\rm las}}(t)=2\pi\nu_{{\rm las}}(t)=\frac{2\pi c}{\lambda_{{\rm las}}(t)}$, $Q_{\rm las}$ is laser pulse amplification factor inside of passive partition of laser resonator, $\Delta\omega_p(t)$ is the laser pulse bandwidth, and $\tau_r^{(j)}$ is random process which determines fluctuations of inverse population build up times from pulse to pulse. As a result, stability of the laser pulse power can be rather poor. To circumvent this problem, optical path length can be also modulated by retro-mirror position, which is controlled by periodic voltage applied to PZT. 
	
	The laser pulse bandwidth evolves according to voltage applied to PZT as following 
	\b
	\Delta\omega_p(t)=\frac{c}{x(t)} \frac{1-\sqrt{RR_W}G_0(t)}{\pi\sqrt{\sqrt{RR_W}G_0(t)}},
	\e 
	where $R_W=R_W(\lambda_{\rm las}(t))=\rho_W^2$ and $R$ are semitransparent window and retromirror power reflectivtity coefficients respectively, $G_0$ is single pass gain. Single pass gain is given by:
	\b 
	G_0=e^{\gamma_0(t)L},
	\e 
	where $L$ is length of the laser cuvette filled by active medium, and emission line shape value at resonance condition is: 
	\b
	\gamma_0(t)=\gamma_0(\nu_0)\frac{\Delta\nu/2}{(\nu_{\rm las}(t)-\nu_0)^2+(\Delta\nu/2)^2},
	\e 
	where $\Delta\nu/2$ is the full width at half of maximum (FWHM) of laser roto-vibrational transition line centered at $\nu_0$. 
	
	As the next step, we take average of Eq.~(\ref{intens}) over the length $W_{IC}$ of the intersection between the laser beam and the gas flow:
	\b
	\label{intens_1}
	\bar I(t)=\frac{c\epsilon_0n}2E_0^2\langle\aleph\tilde\aleph\rangle_x.
	\e
	Here,
	\begin{eqnarray}
		\label{intens_2}
		\aleph(x,t)&=&\sum\limits_{j=0}^{N_p(t)}\Phi^{(j)}(x,t),\qquad \tilde\aleph(x,t)=\sum\limits_{j=0}^{N_p(t)}\tilde\Phi^{(j)}(x,t),\\
		\tilde\Phi^{(j)}(x,t)&=&\frac1{2\pi}\int\limits_{-\infty}^\infty d\omega\overline{Q}_{\rm las}(\omega,x,t){\cal F}(\omega,t){\rm e}^{i\omega t_{j}},
	\end{eqnarray}
	where $\overline{f}$ means complex conjugate and averaging is carried out as \b\label{mean}
	\langle f\rangle_x=\frac{1}{W_{\rm flow}}\int\limits_{\delta}^{\delta+W_{\rm flow}}f(x)dx.
	\e
	Here, $W_{\rm flow}$ is the width of the gas flow and $\delta$ is the distance between the cold core surface of the gas flow and the irradiation cell wall.
	
	Taking into account that
	\b
	\beth(\omega,t):=\sum\limits_{k=0}^{N_p(t)}e^{-ik\omega T_p}=\frac{1-e^{-iN_p(t)\omega T_p}}{1-e^{-i\omega T_p}},
	\e
	quantity $\tilde\aleph$ can be represented as inverse Fourier-transform, \b\label{fourier}\tilde\aleph(x,t)=F^{-1}[\daleth(\omega,x,t)],
	\e
	of the quantity
	\b\ba[c]{l}
	\daleth(\omega,x,t)=\beth(\omega,t){\cal F}(\omega,t)\overline{Q}_{\rm las}(\omega,x,t)e^{-i\omega\tau_j}
	\ea
	\e
	Here $\tau_j=\langle\tau_r^{(j)}\rangle_t$, where the mean value is taken over some observation time interval $t$, that is much larger than time between the laser pulses.
	
	Generalization of the obtained in Ref.~\cite{Lya2016} result for spatial distribution of laser electric field amplification factor $Q_{\rm las}$, caused by multiple laser beam reflections inside the multi-pass cavity in the case of cross-wise intersection with gas flow, to a finite time interval is given by: 
	\b
	\label{ampl} Q_{\rm las}(\omega,x,t)=i{\cal T}\frac{1-(\tilde\rho\tilde\rho_W)^\frac{t}{\Delta t}}{1-\tilde\rho\tilde\rho_W}\l[\tilde\rho e^{i\alpha^{(1)}(x)}{\cal D}+e^{i\alpha^{(0)}(x)}\right],
	\e 
	where $\Delta t=\frac{W_{IC}}{c}$, ${\cal T}$ is semitransparent window electric field transmission coefficient at given wavelength,
	\b
	\tilde\rho_W=\rho_W{\cal D},\qquad\tilde\rho=\rho{\cal D},\qquad {\cal D}=\exp\l(-\frac{\sigma_A(\omega)n_{\rm g}W_{\rm flow}}{2}\right),
	\e 
	where $\rho$ is the mirror strip electric field reflectivity, $\sigma_A(\omega)$ is photo-absorption cross section spectrum, $n_{\rm g}=\frac{\mu P}{k_BT}$ is target gas density dissolved at molar fraction $\mu$ in carrier gas, and electric field phase is given by:
	\be
	\alpha^{(0)}(x)&=&\frac{\omega}{c}x+\frac{\sigma_A(\omega)n_{\rm g}}{2}(x-\delta);\\
	\alpha^{(1)}(x)&=&\frac{\omega}{c}(2W_{IC}-x)+i\frac{\sigma_A(\omega)n_{\rm g}}{2}(W_{IC}-x-\delta).
	\ee
	
	\section{Laser Excitation Rate}\label{Sec:Excitation}
	
	Laser excitation rate which corresponds to the sequence of laser pulses propagating in the irradiation cell implemented as a resonator, can be estimated as:
	\b
	\label{abs_rate_p}
	k_A=\frac1{\pi R_L^2}F^{-1}\l[\frac{d\gimel}{d\omega}\right].
	\e
	Here $R_L$ is laser beam radius and 
	\b
	\frac{d\gimel}{d\omega}=\frac1{\hbar\omega}\frac{dW_A}{d\omega}
	\e
	is the spectral number density of absorbed photons, where related spectral energy density can be represented as
	\b
	\frac{dW_A}{d\omega}=\sigma_A(\omega)\frac{c\epsilon_0n}{2}E_0^2\langle\aleph(x,t)\daleth(\omega,x,t)\rangle_x.
	\e 
	Details of this calculation can be found in \cite{Lya2020_a}. The laser beam radius should be large  enough to excite a reasonable amount of the molecules of the selected target component.
	
	\section{Photo-absorption cross section}\label{Sec:Crosssection}
	
	Generalization of the results of~\cite{Eer76} for photo-absorption cross section at fixed ambient gas temperature $T$ and pressure $P$ taking into account different sorts of molecules contribution to time dependent case is straightforward. Its expression corresponding to one laser pulse of length $\tau_p$ is 
	\b \sigma_{\rm
		A}(\omega)=\frac1{\sqrt{1+I(\tau_p)/I_{\rm
				sat}}}\sum\limits_{k=1}^{n_{\rm is}}x_k(\tau_p)\sum\limits_{h}f_{\rm
		vib}^{(k)}(v_{1h}\ldots v_{n_fh},T)g^{(k)}(\tilde\lambda,v_{1h}\ldots v_{n_fh},P,T),\\
	\e
	where $x_k$ is a relative abundancy of molecules of $k$th sort, $n_{\rm is}$ is the number of sorts of molecules and 
	\begin{eqnarray*}
		f_{\rm vib}^{(k)}&=&\prod\limits_{\beta=1}^{n_{\rm f}}\frac{w_{\beta_m}}{w_{\beta_h}}e^{-\frac{hv_{\beta_h}c\tilde\lambda_\beta}{k_BT}}\l(1-e^{-\frac{hc\tilde\lambda_\beta}{k_BT}}\right),\\
		w_\beta&=&
		\begin{cases}
			1 & {\rm for} ~d_\beta=1,\\
			w_\beta=v_\beta+1 & {\rm for}~d_\beta=2,\\
			w_\beta=\frac12(v_\beta+1)(v_\beta+2) & {\rm for}~d_\beta=3
		\end{cases}\\
		g^{(k)}&=&\Delta\tilde\lambda_{mn}\Bigl[I_{\rm IR}^{(k)}(\tau_p,\nu_{mn;h}^P,0)g_v^P(0)f_v^P(0,v_{1h},\ldots,v_{n_fh})\nonumber\\
		&&+\sum\limits_{J=1}^{\infty}\sum\limits_{a=Q,R}I_{\rm IR}^{(k)}(\tau_p,\nu_{mn;h}^a,J)g_v^a(J)f_v^a(J,v_{1h},\ldots,v_{n_fh})\Bigr],\\
		g_v^P&=&\frac{(2J+3)^2}{6k_BT(1-\xi)}\sqrt{\frac{B}{\pi k_BT}}\exp\l\{-\frac{B(J+1)(J+2\xi)}{k_BT}\right\},\\
		g_v^Q&=&\frac{(2J+1)^2}{6k_BT\xi_B(J+1)}\sqrt{\frac{B}{\pi k_BT}}\exp\left\{-\frac{B(J+1)J}{k_BT}\right\},\\
		g_v^R&=&\frac{(2J-1)^2}{6k_BT(1-\xi)}\sqrt{\frac{B}{\pi k_BT}}\exp\left\{-\frac{B(J+1-2\xi)J}{k_BT}\right\},\\
		\Delta\tilde\lambda_{mn}&=&\frac{2(1-\xi)\sqrt{Bk_BT}}{hc}\, .
	\end{eqnarray*}
	Here $I_{\rm sat}$ is the laser saturation intensity, $n_{\rm f}$ is the number of fundamental vibrational modes, $v_{\beta_h}$ is the
	vibrational number of the mode $\beta_h$, which corresponds to the hot band $h$;
	$B=h\nu_B$, $\xi$, and $\xi_B$ are rotational, Coriolis, and
	vibrational stretch effect on rotation constants respectively. The
	latter can be approximated~\cite{Eer76} as
	$$\xi_B=\frac{\Delta\nu_B}{\nu_B}\approx\frac{\nu_B}{\nu_{\rm max}-\nu_{mn}}(1-\xi)^2,$$
	where $\nu_{\rm max}$ corresponds to
	R-branch high-frequency edge. 
	
	Absorption line FWHM caused solely by collisional broadening is
	\be\label{FWHM_C}
	\Delta\tilde\lambda_{\rm c}^k[cm^{-1}]&=&2\gamma^k=10^{-2}\frac{\Delta\nu_{\rm c}^k}{c},\\
	\Delta\nu_{\rm c}^k&=&2n\l(\sigma_{QG}(1-\mu)\sqrt{\frac{8RT}{\pi M_{QG}^k}}+\sigma_{QQ}\mu\sqrt{\frac{8RT}{\pi M_{QQ}^k}}\right).
	\ee
	Here $n=\frac{P}{k_BT}$, $M_{QQ}^k$ and $M_{QG}^k$ are the reduced
	molar masses for collisions of $k$th component (Q) and carrier
	(G) gas molecules, $\sigma_{QQ}$ and $\sigma_{QG}$ are respective
	cross sections. Absorption FWHM caused by Doppler broadening solely
	is ($a=P,Q,R$ denote rotational transitions, index $P$ corresponds to transition with angular momentum change $\Delta J=-1$, $Q$ to $\Delta J=0$, $R$ to $\Delta J=+1$)
	\be
	\Delta\tilde\lambda_{D;k}^a[cm^{-1}]&=&\sigma_{0;k}^a\sqrt{8\ln2}=10^{-2}\frac{\Delta\nu_{D;k}^a}{c},\qquad a=P,Q,R,\\
	\Delta\nu_{D;k}^a&=&\frac{\nu_{mn;h}^a}{c}\sqrt{\frac{8\ln(2)R_BT}{M_Q^k}},\qquad a=P,Q,R
	\ee
	where $M_Q^k$ is $k$th component molar mass.
	
	In the general case of arbitrary gas flow pressure and temperature, photo-absorption cross section spectral
	density is estimated by Voigt profile, which takes into account interaction of gas flow molecules with laser field standing waves in resonator ($a=P, Q, R$): 
	\[
	f_v^a(\tilde\lambda,v_{1h},\ldots,v_{n_fh},J,P,T)=\frac{2\gamma^k}{\pi\sqrt{2\pi}\sigma_{0;k}^a}
	\int\limits_{-\infty}^\infty\frac{\exp\l\{\frac{-\tilde\lambda^{\prime2}}{2(\sigma_{0;k}^a)^2}\right\}}{(\tilde\lambda-\tilde\lambda_{mn;h}^a-\tilde\lambda^\prime)^2+(\gamma^k)^2} d\tilde\lambda^\prime,
	\]	
	where 
	\be
	\tilde\lambda_{mn;h}^a&=&\frac{\nu_{mn;h}^a}{c}\\
	\nu_{mn;h}^P&=&\nu_{mn}^{(k)}-\Delta_h-\nu_B(2(1-\xi)J+\xi_BJ^2),\\
	\nu_{mn}^{(k)}&=&\sum\limits_\beta(v_{\beta_m}-v_{\beta_n})\nu_\beta^{(k)};\\
	\nu_{mn;h}^Q&=&\nu_{mn}^{(k)}-\Delta_h-\nu_B\xi_B(J^2-J),\\
	\nu_{mn;h}^R&=&\nu_{mn}^{(k)}-\Delta_h+\nu_B(2(1-\xi)J-\xi_BJ^2),\\
	\Delta_h&=&2\sum\limits_\alpha\sum\limits_\beta\sqrt{\nu_\alpha\nu_\beta
		x_\alpha x_\beta}(v_{\alpha;m}-v_{\alpha;n})v_{\beta h}\, .
	\ee
	Here $x_\alpha$ is anharmonicity constant of fundamental mode $\alpha$, index $m$
	corresponds to the final state.

\section{Summary}\label{Sec:Conclusion}
In this paper general method to maximize the discrimination rate between quantum systems with similar spectra in the method of laser assisted retardation of condensation in rather dense medium is proposed. The proposed scheme exploits voltage applied to modulate PZT deformation to subsequently manipulate the laser field in the resonator in which gas mixture of quantum systems is immersed. The control goal is to maximally excite selected component in the mixture while minimally excite all other components. The optimization is based on solving the optimal control problem to find laser pulse shape.  It can be done using gradient ascent search, for which an explicit expression for the gradient of the control objective functional is derived.  The system under control is distributed along the cell and the control pulse along width of the cell. Enhancement of separation rate of molecules of specific sort is expected due to increase of the difference between diffusion rates of light and heavy species in supersonic over cooled gas flow by the optimal choice of the laser field within the resonator. Lagrange multipliers method may also been employed which according to~\cite{Jak90} allows to save more than two orders of magnitude of computational time comparing to direct propagation of equivalent wave packets. In general, a feedback between the output signal and the PZT with learning control~\cite{Ref4} and possible nonlinear reaction term can be used to better manipulate the resonator if some relevant parameters of the components, such as for example some relevant matrix elements of their dipole moments, are unknown. In this case the problem becomes optimization of stationary states of a distributed system with diffusion and non-linear reaction term. 

\acknowledgments
{This research was supported by the Ministry of Science and Higher Education of the Russian Federation 
(project 0718-2020-0025).}

\end{document}